\documentclass[aps,amsmath]{revtex4}
\usepackage{epsf}
\usepackage{amsmath,amsfonts}
\newcommand{\fr}[2]{\frac{#1}{#2}}
\newcommand{\Ref}[1]{(\ref{#1})}
\newcommand{\be}{\begin{equation}}
\newcommand{\ee}{\end{equation}}
\newcommand{\bn}{\begin{eqnarray}}
\newcommand{\en}{\end{eqnarray}}
\newcommand{\bd}{\begin{displaymath}}
\newcommand{\ed}{\end{displaymath}}
\newcommand{\bnn}{\begin{eqnarray*}}
\newcommand{\enn}{\end{eqnarray*}}
\newcommand{\adb}{\allowdisplaybreaks }
\newcommand{\bs}{\begin{subequations}}
\newcommand{\es}{\end{subequations}}
\newcommand{\non}{\nonumber}
\newcommand{\diag}{\textrm{diag}}
\newcommand{\N}{{\not\hspace{-0.8ex}D}}
\begin{document}
%--------------------------------------------------------------------

\title{Vacuum Expectation Value of the Spinor Massive field
in the Cosmic String Space-Time}

\author{Valdir B. Bezerra\footnote{e-mail: valdir@fisica.ufpb.br} and \
Nail R. Khusnutdinov\footnote{e-mail: nail@kazan-spu.ru}
\footnote{On leave from Department of Physics, Tatar State
University of Humanity and Pedagogic, Mezhlauk 1, Kazan 420021,
Russia}}

\address{Departamento de F\'{\i}sica, Universidade Federal da
Para\'{\i}ba, Caixa Postal 5008, CEP 58051-970 Jo\~ao Pessoa, Pb,
Brazil}

%\date{\today}

\begin{abstract}
We found the contribution to the vacuum expectation value of the energy-momentum
tensor of a massive Dirac field due to the conical geometry of the cosmic string
space-time. The heat kernel and heat kernel expansion for the squared Dirac operator
in this background are also considered and the first three coefficients were found in
an explicity form.
\end{abstract}
%\pacs{11.10.Gh; 11.10.-z; 03.70.+k} \keywords{zeta-function;
%Casimir effect; Zero-point energy; Renormalization; Heat-kernel
%coefficients}

\maketitle

%%%%%%%%%%%%%%%%%%%%%%%%%%%%%%%%%%%%%%%%%%%%%%%%%%%%%%%%%%
\section{Introduction}
%%%%%%%%%%%%%%%%%%%%%%%%%%%%%%%%%%%%%%%%%%%%%%%%%%%%%%%%%%
The study of quantum fields in the space-time of a static cylindrically
symmetric cosmic string has been done in different situations. Examples of these
studies are the computation of the non-vanishing contribution to the
vacuum expectation value of the energy-momentum tensor of quantum
fields, as for example, scalar, spinor and vector fields
\cite{HelKon86,Dow87,FroSer87,Matsas,AllLauOtt92,SouSah92%
,SkaHarJas94,GuiLin94,LM,ERBM,ZerCogVan96,AliHorOzd97}. Quantum
fields at non-zero temperature were also considered in the
literature
\cite{DavSah88,Smi89,Lin87,Lin92,CogKirVan94,FroPinZel95,Lin95,Gui95,Lin96b}.
It was noted that the energy-momentum tensor has a non-integrable
singularity at the origin and possesses the structure similar to the
energy-momentum tensor in the wedge  \cite{DeuCan79}. On the other
hand there is a great interest connected with properties of the heat
kernel of the Laplacian like operator and its expansion in this
background. The point to be considered is that in this space-time
there is a singular surface with codimension two which originates
some problems  \cite{KirLoyPar05}. As a consequence, it is
impossible to apply standard expressions for the heat kernel
coefficients as discussed in  \cite{Vas03} and therefore we have to
modify them. Nevertheless, there is a way to calculate the heat
kernel coefficients by considering generalized cone in dimensions
greater then four  \cite{BorKirDow96} and to treat the heat kernel
coefficients in the standard way using dimensional reduction.
General considerations concerning the first heat kernel coefficient
and up to systems with spin-$2$ were considered in Ref.
\cite{FurMie97}.

Cosmic string \cite{7,Vil81} is an exotic topological defect
\cite{9} which may have been formed at phase transitions in the
very early history of the Universe. Up to the moment no direct
observational evidence of their existence has been found (see,
nevertheless \cite{Saz03}), but the richness of the new ideas they
brought along to general relativity seems to justify the interest
in the study of these structures.

The gravitational field of a straight infinitely long cosmic
string is quite remarkable; a particle placed at rest around a
straight, infinite, static string will not be attracted to it;
there is no local gravity. The space-time around a cosmic string
is locally flat but not globally. The external gravitational field
due to a cosmic string may be approximately described by a
commonly called conical geometry. Due to this conical geometry a
cosmic string can induce several effects like, for example,
gravitational lens \cite{Vil81,10}, pair production \cite{11},
electrostatic self-force \cite{12} on an electric charge at rest,
bremsstrahlung process \cite{13,13a} and the so-called
gravitational Aharonov-Bohm effect \cite{14}.

In the present paper we obtain an explicit form of the expression
for vacuum expectation value of the energy-momentum tensor of a
massive Dirac field in the cosmic string space-time, showing up that
the conical structure of this space-time induces a non-vanishing
value for this quantity. The massless case was already considered by
Frolov and Serebriany \cite{FroSer87}. The energy-momentum tensor is
traceless and it has the same non-integrable singularity at the
string as in the massless case. We also reobtain the heat kernel and
heat kernel expansion in terms of contour integrals. We show that in
the coincidence limit, the Green function of squared Dirac operator
is defined by the first heat kernel coefficient and the energy-momentum
tensor is defined by the second one. The vacuum expectation value of
the energy-momentum tensor of a massive scalar field was considered
in Refs. \cite{GuiLin94,Iel97}.

The paper is organized as follows. In Sec. \ref{Sec:Green} we find
the Green function of the squared Dirac operator and the
corresponding heat kernel. In Sec. \ref{Sec:Heat} we obtain the
expansion of the heat kernel and found in manifest form the heat
kernel coefficients which are expressed in terms of the delta function
and its derivatives. The vacuum expectation value for the Dirac
field is obtained in Sec. \ref{Sec:VEV}. We conclude with the
Sec.\ref{Sec:Concl} by summarizing the results obtained presenting
some remarks. In this paper we adopt the units where $\hbar = c = G
=1$.

%%%%%%%%%%%%%%%%%%%%%%%%%%%%%%%%%%%%%%%%%%%%%%%%%%%%%%%%%%
\section{Green Function and Heat Kernel}\label{Sec:Green}
%%%%%%%%%%%%%%%%%%%%%%%%%%%%%%%%%%%%%%%%%%%%%%%%%%%%%%%%%%

To start with let us consider the massive Dirac equation in the
Euclidean sector, $\tau = it$, in the space-time generated by a
cosmic string. The space-time is described by the following line
element  \cite{Vil81}
\bd
 ds^2 = d\tau^2 + d\rho^2 + \fr{\rho^2}{\nu^2}d\varphi^2 + dz^2
\ed
where $\tau,z \in \mathbb{R}, \rho \in \mathbb{R}^+$ and $\varphi
\in [0,2\pi]$. The manifold associated with a cosmic string, ${\cal
M}={\cal \mathbb{R}}^2\otimes {\cal C}_\nu^2$, is a direct product
of two the dimensional space ${\cal \mathbb{R}}^2$ and two
dimensional conical space ${\cal C}_\nu^2$. The scalar curvature has
to be understood as a distribution  \cite{SokSta77}
\be\label{RSing}
\mathcal{R} = 4\pi \fr{\nu -1}\nu\fr{\delta^{(2)}(x)}{\sqrt {g}}.
\ee
The general considerations about the properties of this space-time
may be found in \cite{Che83} while the Green function of the Dirac
equation, for a massive field in this background was found in Ref.
\cite{Lin95}. We also obtain the Green function in this
background, but in different form compared with the one obtained
in \cite{Lin95} which is more suitable for our proposal.

According the general considerations concerning the Dirac equation
in arbitrary dimensions \cite{DowAppKirBor96,Kir01,GroAndCar02},
the Euclidean spinor Green function obeys the following equation
\bd
\N_m {\cal S} (x;x') = -I_4 \fr{\delta^{(4)}(x-x')}{\sqrt{g}},
\ed
where $\N_m = \gamma^\mu \widetilde{\nabla}_\mu + m$ is the Dirac
operator and $\widetilde{\nabla}_\mu = \partial_\mu + \Gamma_\mu$.
The  Fock-Ivanenko spinor connection
\bd
\Gamma_\mu = \fr 12 \sigma^{ab} e_{(a)}^\nu\nabla_\mu e_{(b)\nu}
\ed
is expressed in terms of the vierbein $e_{(b)\nu}$ and the matrix
$\sigma^{ab} = \fr 14 [\gamma^{(a)},\gamma^{(b)}]$. Gamma matrices
satisfy the Clifford commutation relation
$\{\gamma^{(a)},\gamma^{(b)}\} = 2\delta^{ab}I_4$ and the
generalized Dirac matrices, $\gamma^{\mu}(x)$, are defined by the
relation $\gamma^{\mu}(x) = e^\mu_{(a)}\gamma^{(a)}$. In what
follows we will adopt the standard representation
\bd
\gamma^{(0)} = \begin{pmatrix}1&0\\0&-1\end{pmatrix},\
\gamma^{(k)} = - i
\begin{pmatrix}0&\sigma^k\\-\sigma^k&0\end{pmatrix},
\ed
where $\sigma^k$ are the Pauli matrices. As to the vierbein, we
will consider the one used in \cite{DeSJac89}, which are given by
the following expressions
\bnn
e^\mu_{(a)} =
\begin{pmatrix}
1&0&0&0\\
0&\cos\varphi&-\fr\nu\rho \sin\varphi&0\\
0&\sin\varphi&+\fr\nu\rho \cos\varphi&0\\
0&0&0&1\end{pmatrix},
\enn
where rows correspond to the number $(a)$ of the vector
$e^\mu_{(a)}$. Thus, we obtain the results
\bd
\gamma^\tau = \gamma^{(0)},\ \gamma^\rho = \cos\varphi\gamma^{(1)}
+ \sin\varphi\gamma^{(2)},\ \gamma^\varphi = -\fr\nu\rho
\sin\varphi\gamma^{(1)} + \fr\nu\rho\cos\varphi\gamma^{(2)},\
\gamma^z = \gamma^{(3)},
\ed
and
\bd
\Gamma_\mu =
-\fr{\nu-1}{2\nu}\gamma^{(2)}\gamma^{(1)}\delta_{\mu,\varphi} =
i\fr{\nu-1}{2\nu}\Sigma_{3}\delta_{\mu,\varphi},
\ed
where $\Sigma_{3}= -i \gamma^{1} \gamma^{2}.$
Therefore, the spinor Green function ${\cal S}$ obeys the following
equation
\bd
(\gamma^\mu \partial_\mu - \fr{\nu-1}{2\rho} \gamma^\rho +m) {\cal
S}(x;x') = - I_4\fr{\delta^4(x-x')}{\sqrt{g}}.
\ed

Let us define the Green function ${\cal G}$ of the squared Dirac
operator by the relation
\be\label{SandG}
{\cal S}(x;x') = \N_{-m}{\cal G}(x;x').
\ee
It obeys the following equation
\bd
\N^2 {\cal G}(x;x') = - I_4 \fr{\delta^4(x-x')}{\sqrt{g}},
\ed
where
\be\label{DefD^2}
\N^2 = \N_m\N_{-m} = \left(g^{\mu\nu}\partial^2_{\mu\nu} + \fr
1\rho \partial_\rho - \fr{(\nu-1)^2}{4\rho^2} +
i\fr{\nu(\nu-1)}{\rho^2}\Sigma_3
\partial_\varphi - m^2 - \fr 14 \mathcal{R}\right).
\ee
This equation has no the form of the equation for the scalar Green
function. Its general form was presented in \cite{Vas03}. In this
paper we will omit the term with the singular scalar curvature.
The contribution arising from it has been discussed in the
literature\cite{AllLauOtt92,AllOtt90,KayStu91,AllKayOtt96}.

Due to the fact that $\Sigma_3$ is diagonal, the Green function
${\cal G}$ is diagonal, too. In order to obtain the Green function
in explicitly form let us find the full set of bispinors which
satisfies the equation
\be\label{Eq:G}
\N^2 \phi(x) = - \lambda^2 \phi(x).
\ee
We shall numerate the eigenfunctions by eigenvalues of operators
$\widehat{p}_\tau,\ \widehat{p}_z$ and the projection of the full
momentum $\widehat{J}_3$, as
\bnn
\widehat{p}_\tau \phi &=& -i \partial_\tau \phi =
E \phi,\non\\
\widehat{p}_z \phi &=& -i \partial_z \phi =  p_3
\phi,\\
\widehat{J}_3 \phi &=& (\widehat{p}_\varphi + \fr 1{2\nu}
\Sigma_3)\phi = (-i\partial_\varphi + \fr 12 \Sigma_3)\phi = j
\phi,\non
\enn
and define the number $p_\bot$ by relation $\lambda^2 = p_\bot^2 +
E^2 +p_3^2 +m^2$. All these operators commute with each other and
with Dirac and squared Dirac operators. The above set of operators
define the eigenfunction up to four arbitrary constants. Let us
numerate these independent solutions by eigenvalues of matrices
$\Sigma_3$ with eigenvalue $a=\pm 1$ and $\gamma_5 =
\gamma^{(0)}\gamma^{(1)}\gamma^{(2)}\gamma^{(3)}$ with eigenvalue
$b=\pm 1$. Both these operators commute with the above set of
operators. We shall numerate eigenfunctions by numbers
$q=E,p_3,p_\bot, l$ where $E,p_3\in \mathbb{R},p_\bot\in
\mathbb{R}^+,l=0,\pm 1, \ldots (j=l+\fr 12)$ and by $(a,b) = (\pm
1,\pm 1)$. Therefore we have the following four independent
solutions
\bnn
\phi_q^{(+,\pm)} &=& \fr{\sqrt{\nu}}{4 \pi^{3/2}}e^{iE\tau +
ip_3 z + i l \varphi} J_{\beta_-}(p_\bot \rho)w_{(+,\pm)},\adb\\
\phi_q^{(-,\pm)} &=& \fr{\sqrt{\nu}}{4 \pi^{3/2}}e^{iE\tau + ip_3
z + i (l+1) \varphi} J_{\beta_+}(p_\bot \rho)w_{(-,\pm)},\non
\enn
where
\bnn
w_{(+,\pm)} = \begin{pmatrix}1\\0\\\pm 1\\0\end{pmatrix},\
w_{(-,\pm)} = \begin{pmatrix}0\\1\\0\\ \pm 1\end{pmatrix}
\enn
are the eigenfunctions of $\Sigma_3$ and $\gamma_5$. Here
$J_{\mu}(z)$ is the Bessel function with indices $\beta_\mp = |\nu
j \mp \fr 12|$. This full set obeys the following relations of
completeness and orthogonality
\bnn
\sum_{l} \iiint d E d p_3 p_\bot dp_\bot \phi_q^{(a,b)}{}^\dag(x')
\phi_q^{(a,b)}(x) =
\fr{\delta^{(4)}(x-x')}{\sqrt{g}},\adb\\
\int \phi_{q'}^{(a',b')}{}^\dag(x') \phi_{q}^{(a,b)}(x)
\sqrt{g}d^4 x = \delta_{a,a'} \delta_{b,b'}
\delta_{l,l'}\delta(E-E') \delta(p_3-p'_3) \fr{\delta (p_\bot -
p_\bot')}{p_\bot}.\non
\enn
Now we are ready to obtain the Green function in manifest form.
To do this, let us represent the Green function in the following
form
\bn\label{Gorigin}
{\cal G}(x;x') &=& \sum_{a,b}\sum_{l} \iiint d E d p_3 p_\bot
dp_\bot \fr{\phi_q^{(a,b)}(x)\otimes\phi_q^{(a,b)}{}^\dag(x')}
{p_\bot^2 + E^2 + p_3^2 + m^2}\adb\\
&=&\fr\nu{(2\pi)^3}\sum_{l} \iiint d E d p_3 p_\bot dp_\bot
\fr{e^{i E \triangle\tau + i p_3 \triangle z}}{p_\bot^2 + E^2 + p_3^2 + m^2}\non\adb \\
&\times& \diag (J_{\beta_-}(r) J_{\beta_-}(r')
e^{il\triangle\varphi},\ J_{\beta_+}(r) J_{\beta_+}(r')
e^{i(l+1)\triangle \varphi},\ J_{\beta_-}(r) J_{\beta_-}(r')
e^{il\triangle\varphi},\ J_{\beta_+}(r) J_{\beta_+}(r')
e^{i(l+1)\triangle\varphi}),\non
\en
where $r=p_\bot \rho,\ r'=p_\bot \rho'$. By changing the variable
$l\to l-1$ in the second and forth terms, we can write the Green
function in more simple form, as
\bnn
{\cal G}(x;x') &=& \fr\nu{(2\pi)^3}\sum_{l} \iiint d E d p_3
p_\bot dp_\bot \fr{e^{i E \triangle\tau + i p_3 \triangle z +
il\triangle\varphi}}{p_\bot^2 + E^2 + p_3^2 + m^2}\adb \\
&\times& \diag (J_{\xi_+}(r) J_{\xi_+}(r'),\ J_{\xi_-}(r)
J_{\xi_-}(r'),\ J_{\xi_+}(r) J_{\xi_+}(r'),\ J_{\xi_-}(r)
J_{\xi_-}(r')),\non
\enn
where $\xi_\pm = |\nu l \pm \fr{\nu-1}{2}|$.

It is easy to find from this expression the heat kernel of the squared
Dirac operator in Eq. \Ref{DefD^2}. It is given by
\bn\label{hksqdir}
{\cal K}(x;x'|s) &=& \fr\nu{(2\pi)^3}\sum_{l} \iiint d E d p_3
p_\bot dp_\bot e^{i E \triangle\tau + i p_3
\triangle z + il\triangle\varphi -s(p_\bot^2 + E^2 + p_3^2 + m^2)}\non\adb \\
&\times& \diag (J_{\xi_+}(r) J_{\xi_+}(r'),\ J_{\xi_-}(r)
J_{\xi_-}(r'),\ J_{\xi_+}(r) J_{\xi_+}(r'),\ J_{\xi_-}(r)
J_{\xi_-}(r'))\non\adb\\
&=&{\cal K}_m^{(2)}(\triangle\tau,\triangle z|s) {\cal
K}^{(2)}_\nu(\triangle \varphi,\rho,\rho'|s),
\en
where ${\cal K}_m^{(2)}(\triangle\tau,\triangle z|s)$ is the heat
kernel of the operator $\widehat{L}_m = \partial^2_{\tau^2} +
\partial^2_{z^2} - m^2$ and ${\cal
K}^{(2)}_\nu(\triangle \varphi,\rho,\rho'|s)$ is the heat kernel of
the operator
\bd
\widehat{L}_\nu = \partial^2_{\rho^2} + \fr 1\rho
\partial_\rho + \fr{\nu^2}{\rho^2}\partial^2_{\varphi^2} - \fr{(\nu-1)^2}{4\rho^2} +
i\fr{\nu(\nu-1)}{\rho^2}\Sigma_3 \partial_\varphi.
\ed
Integrating over $E,p_3,p_\bot$, we get
\bs\label{hk}
\bn
{\cal K}_m^{(2)}(\triangle\tau,\triangle z|s) &=& \fr 1{4\pi s}
e^{-m^2 s - \fr{\triangle\tau^2 + \triangle z^2}{4s}},\adb\\
{\cal K}^{(2)}_\nu(\triangle \varphi,\rho,\rho'|s) &=& \fr\nu{4\pi
s} \sum_l e^{il\triangle \varphi - \fr{\rho^2 + \rho'^2}{4s}} \diag
(I_{\beta_-}(\fr{\rho\rho'}{2s}),\
I_{\beta_+}(\fr{\rho\rho'}{2s})e^{i\triangle\varphi},\
I_{\beta_-}(\fr{\rho\rho'}{2s}),\
I_{\beta_+}(\fr{\rho\rho'}{2s})e^{i\triangle\varphi})\adb\non\\
&=& \fr\nu{4\pi s} \sum_{l}e^{il\triangle \varphi - \fr{\rho^2 +
\rho'^2}{4s}} \diag (I_{\xi_+}(\fr{\rho\rho'}{2s}),\
I_{\xi_-}(\fr{\rho\rho'}{2s}),\ I_{\xi_+}(\fr{\rho\rho'}{2s}),\
I_{\xi_-}(\fr{\rho\rho'}{2s})). \label{hkDiracSquare}
\en
It is worth noting that this expression leads to a heat
kernel which is a diagonal matrix. The components of this matrix are
defined in terms of the first component ${\cal K}_{1,1}$ which we
denote by $\mathbf{K}$ by relations: ${\cal K}_{3,3} = \mathbf{K}$
and ${\cal K}_{2,2} = {\cal K}_{4,4}= \mathbf{K}^*$. Indeed,
because $\xi_{\pm}\to \xi_{\mp}$ with changing $l \to \-l$ then
\bd
\sum_{l} e^{il\triangle\varphi}I_{\xi_-} = \sum_{l}
e^{-il\triangle\varphi}I_{\xi_+} = \left(\sum_{l}
e^{il\triangle\varphi}I_{\xi_+} \right)^*,
\ed
and the heat kernel has the following structure
\be\label{hkDiracSquareStructure}
{\cal K}^{(2)}_{\nu} =
\diag\left(\mathbf{K},\mathbf{K}^*,\mathbf{K},\mathbf{K}^*\right).
\ee
\es

Using Eqs. \Ref{SandG} and \Ref{hksqdir} we obtain the spinor
Green function in manifest form, which can be written as
\be\label{SEucl}
\mathcal{S}(x;x') = \fr\nu{(4\pi)^2} \sum_{l} e^{ il\triangle
\varphi}\int_0^\infty \fr{ds}{s^2} e^{- s m^2 - \fr{\triangle\tau^2
+ \triangle z^2 + \rho^2 + \rho'^2}{4s}} (-m S_1 -
\fr{\triangle\tau}{2s} S_2 + i\fr{\triangle\ z}{2s} S_3 - \fr{i}{2s}
S_4)
\ee
where
\bn
S_1 &=& \diag \left(I_{\beta_-}, I_{\beta_+}e^{i\triangle\varphi},
I_{\beta_-}, I_{\beta_+}e^{i\triangle\varphi} \right), \  S_2 =
\diag \left(I_{\beta_-}, I_{\beta_+}e^{i\triangle\varphi},
-I_{\beta_-}, -I_{\beta_+}e^{i\triangle\varphi} \right),\adb\non\\
S_3 &=& \left(%
\begin{array}{cccc}
  0 & 0 & I_{\beta_-} & 0 \\
  0 & 0 & 0 & -I_{\beta_+}e^{i\triangle\varphi} \\
  -I_{\beta_-} & 0 & 0 & 0 \\
  0 & I_{\beta_+}e^{i\triangle\varphi} & 0 & 0 \\
\end{array}%
\right),\adb\\
S_4 &=& \left(%
\begin{array}{cccc}
  0 & 0 & 0 & e^{-i\varphi'} (\rho' I_{\beta_-} - \rho I_{\beta_+})\\
  0 & 0 & e^{i\varphi} (\rho' I_{\beta_+} - \rho I_{\beta_-}) & 0 \\
  0 & -e^{-i\varphi'} (\rho' I_{\beta_-} - \rho I_{\beta_+}) & 0 & 0 \\
  -e^{i\varphi} (\rho' I_{\beta_+} - \rho I_{\beta_-}) & 0 & 0 & 0 \\
\end{array}%
\right)\non
\en
and the argument of the Bessel functions of second kind
$I_{\beta_\pm}$ is $z=\fr{\rho\rho'}{2s}$. To obtain previous
expression we used the relations :
\bd
I'_{\beta_-} - \fr{\epsilon_l\beta_-}{z} I_{\beta_-} = I_{\beta_+},\
I'_{\beta_+} + \fr{\epsilon_l\beta_+}{z} I_{\beta_+} = I_{\beta_-},
\ed
where $\epsilon_l = \mathrm{sgn} (l)$.

Now, let us consider some limiting cases. In the massless case,
$m=0$, the Green function can be found in a closed form and is given
explicitly by
\bd
\mathcal{G}(x;x') = \fr\nu{8\pi^2} \fr 1{\rho\rho'}
\fr{1}{\sinh\eta} \fr{\sinh(\fr{\nu+1}2\eta) I_4 + \sinh
(\fr{\nu-1}2\eta) e^{-i \triangle\varphi \Sigma_3}}{\cosh\nu\eta -
\cos\triangle\varphi}.
\ed
We note that this Green function does not coincide with that
obtained in \cite{FroSer87,Lin95} and it does not obey the
relation
\bd
\mathcal{G}(\varphi + 2\pi) = - \mathcal{G}(\varphi).
\ed
Instead of this, it obeys the relation above with minus sign
changed to plus sign. The reason for this fact is connected with
the use of another vierbein as compared with the one used in
\cite{FroSer87}.

In order to reobtain the result showed in \cite{FroSer87}, we have
to apply a rotation of the vierbein we have chosen by an angle
$\varphi$, in the plane $(\rho, \varphi)$. It is well-known that
the bispinor will be multiplied by a factor $e^{\fr i2 \varphi
\Sigma_3}$. The Green function, as a bispinor at point $x$ and
point $x'$, is multiplied by factor $e^{\fr i2\triangle \varphi
\Sigma_3}$. Thus, multiplying by this factor we arrive at the
formula
\bd
\mathcal{G}(x;x') = \fr\nu{8\pi^2} \fr 1{\rho\rho'}
\fr{1}{\sinh\eta} \fr{e^{\fr i2\triangle \varphi \Sigma_3}
\sinh(\fr{\nu+1}2\eta) + e^{-\fr i2\triangle \varphi \Sigma_3}
\sinh (\fr{\nu-1}2\eta)}{\cosh\nu\eta - \cos\triangle\varphi}
\ed
which coincides with the results obtained in
\cite{FroSer87,Lin95}, if we take into account that $i\Sigma_3 =
\gamma^{(1)} \gamma^{(2)}$.

In the case of zero angle deficit, $\nu = 1$, we have $\xi_\pm = |l|$ and thus
\bd
\mathcal{K}(x;x'|s) = \fr 1{(4\pi s)^2} e^{-m^2 s -
\fr{R^2}{4s}}I_4,
\ed
where $R =(\triangle\tau^2 + \triangle z^2 + \rho^2 + \rho'^2 - 2
\rho\rho'\cos\triangle\varphi)^{1/2}$ is the distance between points
$x$ and $x'$ in the Euclidean 4-dimensional space in cylindrical
coordinates. The Green function in this case turns into
\bd
\mathcal{G}(x;x') = \int_0^\infty \mathcal{K}(x;x'|s) ds = \fr
1{4\pi^2}\fr mR K_1(mR)I_4,
\ed
as it should be. Here $K_\mu$ is the Bessel function of
second kind.

%%%%%%%%%%%%%%%%%%%%%%%%%%%%%%%%%%%%%%%%%%%%%%%%
\section{Heat Kernel Expansion}\label{Sec:Heat}
%%%%%%%%%%%%%%%%%%%%%%%%%%%%%%%%%%%%%%%%%%%%%%%%%%%%%%%

Let us investigate the asymptotic expansion of the heat kernel of
the squared Dirac operator which is given by Eqs.
\Ref{hkDiracSquare} and \Ref{hkDiracSquareStructure}. It is
impossible to take an expansion of the Bessel function because the
asymptotic expansion of the Bessel function depends on the ratio
argument and indices. For this reason we use the integral
representation for the Bessel function, as follows
\bd
I_\mu (z) = \fr{1}{2\pi} \int_{\Gamma} e^{i(w-\fr\pi 2)\mu + z\sin
w} dw.
\ed
The contour $\Gamma$ lies in the half-strip $\Re z \in [-\fr\pi
2,\fr{3\pi}2], \Im z >0$. It goes from $-\fr\pi 2 + i\infty$ to
$\fr{3\pi}2 + i\infty$.

Let us consider, in general, the series
\bd
f(\alpha,r)=\nu\sum_{l}e^{il\triangle\varphi} I_{|\nu (l +
\alpha)|}(r),
\ed
where $|\alpha| \leq 1$. For our cases we have $\alpha = \pm
\fr{\nu -1}{2\nu}$ and $|\alpha| < \fr 12$. Then we use the
representation above for the Bessel function and change the
integration and summation and shift the variable $w$ to $z= w-\fr\pi 2$.
In this situation, we obtain the following expression for the
function $f(\alpha,r)$
\bd
f(\alpha,r) = \fr\nu{2\pi}\int_{\Gamma'} e^{r \cos z} \left[
\fr{e^{-iz|\nu\alpha|}}{e^{-i(z\nu -
\epsilon_\alpha\triangle\varphi)}-1} -
\fr{e^{iz|\nu\alpha|}}{e^{i(z \nu + \epsilon_\alpha
\triangle\varphi)}-1}\right]dz,
\ed
where $\epsilon_\alpha = sgn(\alpha)$. The contour $\Gamma'$ is
the contour $\Gamma$ shifted by $\pi/2$ to the left. It lies in
the positive part of the strip $z\in [-\pi,\pi]$. Then we divide
the integral in two parts according with integrand and in the
second part, we do the change of variable $z\to -z$. The integrand takes
the form as the first integrand but the contour is the central
symmetry of $\Gamma'$. Therefore we may recombine both integrals into
a single integral and arrive at the following formula for function
$f(\alpha,r)$
\bd
f(\alpha,r) = \fr\nu{2\pi}\int_{\gamma} \fr{e^{-iz|\nu\alpha| + r
\cos z}}{e^{-i(z\nu - \epsilon_\alpha\triangle\varphi)} - 1}dz.
\ed
The contour $\gamma$ has two branches. First one is contour
$\Gamma'$ and the second is the central symmetry of $\Gamma'$.
Taking into account this formula we arrive at the following
expressions for heat kernel component
\bd
\mathbf{K} = \fr{e^{-\fr{\rho^2 + \rho'^2}{4s}}}{4\pi s}
f(\fr{\nu-1}{2\nu},\fr{\rho\rho'}{2s}) = \fr{\nu e^{-\fr{\rho^2 +
\rho'^2}{4s}}}{8\pi^2 s} \int_{\gamma} \fr{e^{-iz \fr{\nu-1}2 +
\fr{\rho\rho'}{2s}\cos z}}{e^{-i(z\nu - \triangle\varphi)} - 1}dz.
\ed
The complex conjugate $\mathbf{K}^*$ differ from $\mathbf{K}$
by the sign of $\triangle\varphi$ only.

Now, let us return to the function $f(\alpha,r)$ and modify the
contour $\gamma$. It is easy to see that the zeros of the
denominator are in the points $z_n = \epsilon_\alpha
\fr{\triangle\varphi}\nu + \fr{2\pi n}{\nu}$. Because of
$|\triangle\varphi | < \pi$ and $\nu > 1$, then we have
that$|\triangle\varphi /\nu| < \pi$. Therefore, the first zero of
the denominator, $z_0 = \epsilon_\alpha \fr{\triangle\varphi}\nu$,
belongs to the interval of integration $[-\pi,\pi]$. Let us
extract this first pole in manifest form. With this we add the
integrals over two lines around point $z=
\epsilon_\alpha\triangle\varphi/\nu$ in each of which we integrate
in opposite directions. The contribution from these additional
integrals are zero. Then we divide the contour $\gamma$ with
additional lines in two parts and represent the integral as an
integral over contour $\chi$ and over a circle around point
$\epsilon_\alpha \fr{\triangle\varphi}\nu$. The contour $\chi$ has
two branches. The first one goes from $-\pi + i\infty$ to $-\pi -
i\infty$, very close to the point $\epsilon_\alpha
\fr{\triangle\varphi}\nu$ from the left side. The second part of the
contour goes from $\pi + i\infty$ to $\pi - i\infty$, very close
to the point $\epsilon_\alpha \fr{\triangle\varphi}\nu$ from the right
side. The second integral is minus the residue at point
$\epsilon_\alpha \fr{\triangle\varphi}\nu$. Therefore we may
represent our function $f(\alpha,r)$ in the form below
\bd
f(\alpha,r) =  e^{r\cos\fr{\triangle\varphi}\nu - i\alpha
\triangle\varphi} + \fr\nu{2\pi} \int_{\chi} \fr{e^{-iz|\nu\alpha|
+ r \cos z}}{e^{-i(z\nu - \epsilon_\alpha\triangle\varphi)}-1}dz.
\ed
Taking into account this representation we arrive at the following
expression for component $\mathbf{K}$
\bd
\mathbf{K} = \fr{e^{-\fr{d^2}{4s} -
i\fr{\nu-1}{2\nu}\triangle\varphi}}{4\pi s}\left[ 1 +
\fr\nu{2\pi}\int_{\chi} \fr{e^{-i(z\nu - \triangle\varphi)
\fr{\nu-1}{2\nu} + \fr{\rho\rho'}{2s}(\cos z -
\cos\fr{\triangle\varphi}\nu)}}{e^{-i(z\nu -
\triangle\varphi)}-1}dz\right],
\ed
where $d^2 = \rho^2 + \rho'^2 - 2\rho\rho' \cos
\fr{\triangle\varphi}\nu$. The expression for the complex conjugate
$\mathbf{K}^*$ component differs from the above by the sign of
$\triangle\varphi$.

Therefore, in the coincidence limit $\rho' = \rho$ and $\varphi' =
\varphi$ we obtain the following expression for the heat kernel
\be\label{K2}
{\cal K}_{\nu}^{(2)}(x;x|s) = \fr{1}{4\pi s}\left[ 1 +
\fr\nu{2\pi}\int_{\chi} \fr{e^{-iz\fr{\nu-1}{2} -
\fr{\rho^2}{s}\sin^2 \fr z2}}{e^{-iz\nu}-1}dz\right]I_4,
\ee
where the contour $\chi$ intersects the real axis close to the
origin. In order to find the expansion of the heat kernel over $s$ we
use the formula presented in  \cite{CogKirVan94}, which is given
by
\bd
\fr{\nu\sigma}{\pi} e^{-\sigma \rho^2} = \sum_{n=0}^{\infty}
\fr{\triangle_{(2)}^n \delta^{(2)}(\vec{r})}{n! (4\sigma)^n},
\ed
where $\delta^{(2)}(\vec{r}) = \fr\nu\rho\delta(\rho) \delta
(\varphi)$ and $\triangle_{(2)} = \partial^2_{\rho^2} + \fr
1\rho\partial_\rho + \fr{\nu^2}{\rho^2} \partial^2_{\varphi^2}$.
The integral may be calculated by the method of residue at the
zero point. Thus, we obtain
\bd
{\cal K}_{\nu}^{(2)}(x;x|s) = \fr{1}{4\pi s}\left[I_4 +
\sum_{n=1}^\infty a_n(x;x) s^n\right],
\ed
where
\bd
a_n(x;x) = I_4\fr{R_{n-1}}{(n-1)! 4^{n-1}}\triangle_{(2)}^{n-1}
\delta^{(2)}(\vec{r})
\ed
and
\bd
R_n = \mathrm{Res}_0 \fr{i\pi e^{-iz \fr{\nu-1}{2}}}{
(e^{-iz\nu}-1)\sin^{2n+2} \fr z2}.
\ed
Therefore, the first three coefficients are given by
\bd
R_0 = -\fr{\pi(\nu^2 - 1)}{6\nu},\ R_1 = \fr{7\nu^2 + 17}{60}
R_0,\ R_2 = \fr{31\nu^4 + 178\nu^2 + 367}{2520}R_0.
\ed
The expansion of the four dimensional heat kernel is expressed in
terms of the same coefficients, as
\bd
{\cal K}(x;x'|s) = \fr{1}{(4\pi s)^2}\left[ I_4 +
\sum_{n=1}^\infty a_n(x;x) s^n\right].
\ed
From the above expression we observe that the first coefficient
$R_0$ for the spinor field is minus $1/2$ of the same coefficient
corresponding to the scalar field. Therefore, we have
\bd
\mathrm{tr} a_1^{(1/2)} = -\fr{N}2 a_1^{(0)} = - 2 a_1^{(0)},
\ed
and thus in accordance with \cite{FurMie97,Fur94,IelMor99}. The
other coefficients have different structure.

%%%%%%%%%%%%%%%%%%%%%%%%%%%%%%%%%%%%%%%%%%%%%%%%
\section{Vacuum Expectation Values}\label{Sec:VEV}
%%%%%%%%%%%%%%%%%%%%%%%%%%%%%%%%%%%%%%%%%%%%%%%%

First of all let us find in closed form the coincidence limit of
the renormalized Green function of the squared Dirac operator. To do the process of
renormalization we subtract the same function with $\nu=1$.
Because the second integral in Eq. \Ref{K2} is zero for $\nu=1$, we
have the following expression in the coincidence limit
\be\label{GCoin}
{\cal G}^{ren}(x;x) = \fr{\nu}{32\pi^3} \int_0^\infty \fr{ds}{
s^2}\int_{\chi} \fr{e^{-iz\fr{\nu-1}{2} - \fr{\rho^2}{s}\sin^2 \fr
z2 - m^2 s}}{e^{-iz\nu}-1}dz I_4.
\ee
Firstly, let us consider the massless case. We may integrate easily
with respect to $s$ and obtain
\bd
{\cal G}^{ren}(x;x) = \fr{\nu}{32\pi^3\rho^2} \int_{\chi}
\fr{e^{-iz\fr{\nu-1}{2}}}{(e^{-iz\nu}-1)\sin^2 \fr z2}dz I_4.
\ed
Then we close the contour to infinities and calculate this expression
with the  residue at the point zero. Thus, we get the result
\be\label{G}
{\cal G}^{ren}(x;x) = \fr{\nu R_0}{16\pi^3\rho^2} I_4 = -
\fr{\nu^2 -1}{96\pi^2\rho^2}I_4.
\ee

\begin{figure}[ht]
\begin{center}
\epsfxsize=9truecm\epsfbox{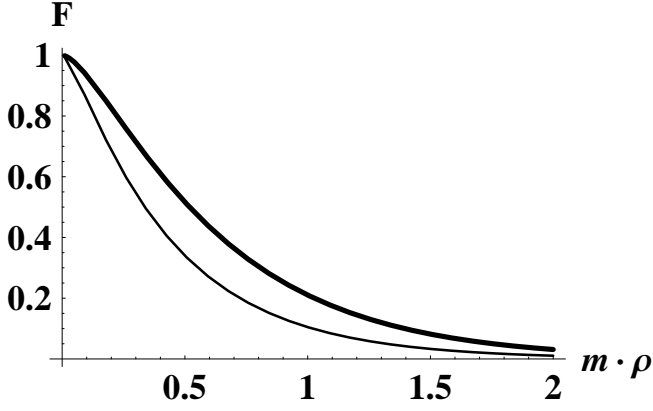}
\end{center} \caption{The plot of the
ratio $F= {\cal G}^{ren}(x;x)_{m\not = 0}/{\cal G}^{ren}(x;x)_{m =
0}$ for $\nu = 0.01$ (thin curve) and $\nu = 2$ (thick curve).}
\label{Fig:f}
\end{figure}

In the massive case, we deform the contour in two straight lines over
$z=\pm \pi$. We have to take into account the residues at points
$z_n = 2\pi n /\nu$, where $|n|<\nu/2$. For $\nu <2$ there is no
zeros in the denominator and no residues. By changing the variable
$z=\pm \pi + i y$ we obtain, in general, the result
\bd
\int_{\chi} \fr{e^{-iz\fr{\nu\pm 1}{2} - \fr{\rho^2}{s}\sin^2 \fr
z2}}{e^{-iz\nu}-1}dz = 4\cos\fr{\pi\nu}2\int_{0}^\infty e^{-
\fr{\rho^2}{s}\cosh^2 \fr y2 } \fr{\sinh
\fr{y}{2}\sinh\fr{y\nu}2}{\cosh \nu y - \cos\pi\nu}dy.
\ed
Taking into account this formula we arrive at the following
expression
\bd
{\cal G}^{ren}(x;x) = \fr {m}{16\pi^2\rho} \sum_{n=1}^{[\nu/2]}
(-1)^n \tan \fr{\pi n}\nu K_1 (2m\rho \sin \fr{\pi n}\nu ) +
\fr{m\nu\cos\fr{\pi\nu}2}{4\pi^3\rho} \int_{0}^\infty
\fr{K_1(2m\rho\cosh \fr y2)}{\cosh \fr y2}
\fr{\sinh\fr{y}{2}\sinh\fr{y\nu}2}{\cosh \nu y - \cos\pi\nu}dy
I_4.
\ed
In the case $\nu<2$, only the last term survives. The appearance
of additional terms, when $\nu>2$, is related with the fact that
additional closed  geodesics in $\mathcal{C}^2_\nu$, appears in
this case \cite{Khu95}. For $m\rho \gg 1$ the above expression
exponentially falls down, according to
\bd
{\cal G}^{ren}(x;x) \approx \fr{\nu R_0}{16\pi^{5/2}}
\fr{m^2e^{-2m\rho}}{(m\rho)^{3/2}}.
\ed
In the opposite case, if $m\rho\ll 1$, the expression for ${\cal
G}^{ren}(x;x)$ is given by Eq. \Ref{G}. The plot of the ratio $F=
{\cal G}^{ren}(x;x)_{m\not = 0}/{\cal G}^{ren}(x;x)_{m = 0}$ is
reproduced in Fig. \ref{Fig:f} as a function of $m\rho$.

To calculate the energy-momentum tensor we use the following
formula \cite{GroAndCar02}
\be\label{Tmunu}
\langle T_{\mu\nu} \rangle^{ren} = -\fr 14 \lim_{x'\to x}
\mathrm{Im}\{ \mathrm{Tr}
[\gamma_{(\mu}^L(\nabla_{\nu)}[\mathcal{S}+\mathcal{S}^c]^{ren} -
g_{\nu)}^{\cdot\lambda'}\nabla_{\lambda'}[\mathcal{S}+\mathcal{S}^c]^{ren})I(x';x)]\},
\ee
where $\mathcal{S}^c$ is the charge conjugate spinor Green function
$\mathcal{S}^c = - C\gamma_L^{(0)T}\mathcal{S}^* \gamma^{(0)}_L
C^\dag  = -\gamma_L^{(2)}\mathcal{S}^* \gamma^{(2)}_L$. In above
formulas $\gamma_L^{\mu}$ are the gamma matrices in Lorentz regime:
$\gamma^{(0)} = \gamma^{(0)}_L,\ \gamma^{(k)} = - i\gamma^{(k)}_L$,
and $\tau = it$. In manifest form
\be
\mathcal{S}^c(x;x') = \fr\nu{(4\pi)^2} \sum_{l} e^{- il\triangle
\varphi}\int_0^\infty \fr{ds}{s^2} e^{- s m^2 - \fr{\triangle\tau^2
+ \triangle z^2 + \rho^2 + \rho'^2}{4s}} (-m S_1^{c} -
\fr{\triangle\tau^*}{2s} S_2^{c} + i\fr{\triangle\ z}{2s} S_3^{c} -
\fr{i}{2s} S_4^{c})
\ee
where
\bn
S_1^{c} &=& \diag \left(I_{\beta_+}e^{-i\triangle\varphi},
I_{\beta_-}, I_{\beta_+}e^{-i\triangle\varphi}, I_{\beta_-}\right),
\ S_2^{c} = \diag \left(-I_{\beta_+}e^{-i\triangle\varphi},
-I_{\beta_-}, I_{\beta_+}e^{-i\triangle\varphi}, I_{\beta_-}\right),\adb\non\\
S_3^{c} &=& \left(%
\begin{array}{cccc}
  0 & 0 & I_{\beta_+}e^{-i\triangle\varphi} & 0 \\
  0 & 0 & 0 &  -I_{\beta_-}\\
  -I_{\beta_+}e^{-i\triangle\varphi} & 0 & 0 & 0 \\
  0 & I_{\beta_-} & 0 & 0 \\
\end{array}%
\right),\adb\\
S_4^{c} &=& \left(%
\begin{array}{cccc}
  0 & 0 & 0 & e^{-i\varphi} (\rho' I_{\beta_+} - \rho I_{\beta_-})\\
  0 & 0 & e^{i\varphi'} (\rho' I_{\beta_-} - \rho I_{\beta_+}) & 0 \\
  0 & -e^{-i\varphi} (\rho' I_{\beta_+} - \rho I_{\beta_-}) & 0 & 0 \\
  -e^{i\varphi'} (\rho' I_{\beta_-} - \rho I_{\beta_+}) & 0 & 0 & 0 \\
\end{array}%
\right)\non
\en
and the argument of the Bessel functions is $z =
\fr{\rho\rho'}{2s}$. The manifest form of the spinor Green function
$\mathcal{S}$ is given by Eq. \Ref{SEucl}. In order to renormalize the
spinor Green functions $\mathcal{S}$ and $\mathcal{S}^c$ we have to
subtract from them the same expression with $\nu=1$. It means that
the renormalized Green functions have the same form in which we have to
make the changes
\bd
\nu I_{\beta_-} \to \nu I_{\beta_-} - I_{|l|},\ \nu I_{\beta_+} \to
\nu I_{\beta_+} - I_{|l+1|}.
\ed
In what follows we assume these changes in the Green functions.
Because the renormalized Green functions are finite in the
coincidence limit we may use $\delta_\nu^{\lambda'}$ instead of
$g_\nu^{\ \lambda'}$. To obtain the energy-momentum tensor we use
the following relations
\bs
\bn
\sum_l I_{\beta_-} &=& \sum_l I_{\beta_+},\label{relationI}\\
\sum_l l I_{\beta_-} &=& - \sum_l (l+1)
I_{\beta_+},\label{relationII}
\en
\es
which can be easy proved by changing $l\to -l-1$ in the sum in the left
hand side. Let us briefly show the main steps of calculations (for
simplicity we consider the case $\nu<2$).

From Eq. \Ref{Tmunu} we conclude that the energy-momentum tensor has
diagonal form with components (no summation):
\bd
T_\mu^\mu = -\fr 14 \lim_{x'^{\mu}\to x^\mu} \mathrm{Im}\
\mathrm{Tr} \gamma_L^{\mu} (\partial_\mu -
\partial_{\mu'})(\mathcal{S} + \mathcal{S}^c)^{ren}.
\ed
In this formula the coincidence limit for all variables except
$x^\mu$ has already done. The contributions from $\mathcal{S}^c$
coincides with contribution from $\mathcal{S}$. Let us consider
consequently different components of the energy-momentum tensor.

$T_t^t$: There is non-zero contribution only due to $S_2$, because
$S_1$ and $S_4$ do not depend on the time:
\bd
T_t^t = \fr{1}{8\pi^2} \int_0^\infty \fr{ds}{s^3} e^{-sm^2}
\left\{e^{-\fr{\rho^2}{2s}} \nu\sum_l
I_{\beta_-}(\fr{\rho^2}{2s})\right\}^{ren}.
\ed

$T_z^z$: The calculations of the $T_z^z$ component is similar. The
non-zero contribution is only due to $S_3$ and we obtain $T_z^z =
T_t^t$, as should be the case due to boost-invariance along
$z$-axis.

$T_\rho^\rho$: In the coincidence limit $S_1$ and $S_4$ survive.
The term $S_1$ gives no contribution because
\bd
\lim_{\rho'\to\rho} (\partial_{\rho} -
\partial_{\rho'})e^{-\fr{\rho^2 + \rho'^2}{4s}}I_{\beta_\pm}(\fr{\rho\rho'}{2s}) =
0.
\ed
The non-zero contribution appears only due to $S_4$. By using the
relations
\bnn
\lim_{\rho'\to\rho} (\partial_{\rho} - \partial_{\rho'})
\left\{\rho' I_{\beta_\mp}(\fr{\rho\rho'}{2s}) - \rho
I_{\beta_\pm}(\fr{\rho\rho'}{2s})\right\} &=& -
I_{\beta_-}(\fr{\rho^2}{2s}) - I_{\beta_+}(\fr{\rho^2}{2s})
\enn
and \Ref{relationI} one has
\bd
\lim_{\rho'\to\rho} \gamma^\rho_L(\partial_{\rho} -
\partial_{\rho'})S_4 = -2 I_{\beta_-}(\fr{\rho^2}{2s})
\gamma^\rho_L{}^2 = 2 I_{\beta_-}(\fr{\rho^2}{2s})I_4.
\ed
Therefore, we obtain $T_\rho^\rho = T_t^t$.

$T_\varphi^\varphi$: In the coincidence limit $S_1$ and $S_4$
survive. The component $S_1$ has no contribution to
$T_\varphi^\varphi$. Indeed, by using Eq. \Ref{relationII} one has
\bd
\mathrm{Tr}\lim_{\varphi'\to \varphi}\gamma_L^\varphi
(\partial_{\varphi} - \partial_{\varphi'}) e^{il\triangle\varphi}
S_1 = 2il I_{\beta_-}(\fr{\rho^2}{2s}) \mathrm{Tr}
\gamma_L^\varphi\diag (1,-1,1,-1) =0.
\ed
The non-zero contribution exists only due to $S_4$. One has
\bnn
\lim_{\varphi'\to \varphi}(\partial_{\varphi} -
\partial_{\varphi'}) e^{il\triangle\varphi} S_4 &=& 2il \lim_{\varphi'\to \varphi} S_4 +
\lim_{\varphi'\to \varphi}(\partial_{\varphi} -
\partial_{\varphi'}) S_4\\
&=& -2l(I_{\beta_-} - I_{\beta_+})\fr{\rho^2}\nu \gamma_L^\varphi -
(I_{\beta_-} - I_{\beta_+})\fr{\rho^2}\nu \gamma_L^\varphi.
\enn
As a consequence of the Eq. \Ref{relationI}, the last term gives no
contribution to $T_\varphi^\varphi$ and we arrive at the expression
\bd
T_\varphi^\varphi = \fr{1}{8\pi^2} \int_0^\infty \fr{ds}{s^3}
e^{-sm^2} \left\{e^{-\fr{\rho^2}{2s}} \nu^2\sum_l l\left[
I_{\beta_-}(\fr{\rho^2}{2s})- I_{\beta_+}(\fr{\rho^2}{2s})\right]
\right\}^{ren}
\ed

Thus, using the results of the last section we obtain
\bd
T^t_t = T_0,\ T^\varphi_\varphi = -3 T_0+ T_1.
\ed
Therefore one has the following structure of the energy-momentum
tensor
\bs\label{TmnGen}
\be\label{Tmn}
\langle T^\mu_\nu \rangle^{ren} = T_0 \diag(1,1,-3,1) +
T_1\diag(0,0,1,0),
\ee
where
\bn
T_0 &=& \fr\nu{16\pi^3}\int_0^\infty \fr{ds}{s^3} e^{-sm^2}
\int_\chi \fr{e^{-iz \fr{\nu-1}2 - \fr{\rho^2}s \sin^2\fr
z2}}{e^{-i\nu z} -1}dz,\\
T_1 &=& -\fr{\nu m^3\cos \fr{\pi\nu}2}{\pi^3\rho} \int_0^\infty
\fr{K_1 (2m\rho\cosh \fr y2)}{\cosh \fr y2} \fr{\sinh \fr y2 \sinh
\fr{\nu y}2}{\cos\nu y - \cos\pi\nu}dy.
\en
In the massless case, $m=0$, we get
\bn\label{TSmalr}
T_0 &=& \fr {\nu R_1}{8\pi^3 \rho^4} = -\fr{(\nu^2 -1)
(7\nu^2 + 17)}{2880 \pi^2 \rho^4},\\
T_1 &=& 0,\non
\en
and in the  massive one, we have
\be
T_0 = \fr{\nu m^2\cos \fr{\pi\nu}2}{2\pi^3\rho^2} \int_0^\infty
\fr{K_2 (2m\rho\cosh \fr y2)}{\cosh^2 \fr y2} \fr{\sinh \fr y2
\sinh \fr{\nu y}2}{\cos\nu y - \cos\pi\nu}dy.
\ee
\es

The energy-momentum tensor obtained obeys the conservation relation:
\bd
\nabla_\mu T^\mu_\nu = 0.
\ed
Indeed, the non-trivial equation for $\nu=r$ reads as
\bd
\partial_r T^r_r + \fr 1r (T^r_r - T_\varphi^\varphi) = 0.
\ed
Taking into account the manifest structure of the energy-momentum
tensor given by \Ref{TmnGen}, above relations takes the following form
\bd
\partial_r T_0 + \fr 1r (4T_0 - T_1) = 0.
\ed
It is easy to verify that this equation is valid due to well-known
relation:
\bd
K'_2(z) = -\fr 2z K_2 - K_1.
\ed

The term $T_1$ appears due to the non-zero mass of the field. It may
be represented by the following form
\bd
T_1 = -m^2 \mathrm{Tr}[\mathcal{G}^{ren}(x;x)] = m
\mathrm{Tr}[\mathcal{S}^{ren}(x;x)],
\ed
and gives right the trace of the energy-momentum tensor
\bd
\langle T^\mu_\mu \rangle^{ren} = T_1.
\ed
For $m\rho \gg 1$, the energy-momentum tensor is exponentially
small
\bd
T_0 \approx \fr{\nu R_1}{8\pi^{5/2}} \fr{m^4
e^{-2m\rho}}{(m\rho)^{5/2}}.
\ed
Therefore the energy is localized very close to the string in a
radius smaller than the Compton length of the spinor particle,
$\rho < m^{-1}$. For $m\rho\ll 1$ the expression for $T_k$ is
given by \Ref{TSmalr}. The  plot of the ratio
$T_{0(m\not=0)}/T_{0(m= 0)}$ looks very similar to the plot in
Fig.\ref{Fig:f}.

%%%%%%%%%%%%%%%%%%%%%%%%%%%%%%%%%%%%%%%%%%%%%%%%
\section{Conclusion}\label{Sec:Concl}
%%%%%%%%%%%%%%%%%%%%%%%%%%%%%%%%%%%%%%%%%%%%%%%%
In this paper we considered the Dirac field in the space-time of
an infinitely thin and straight cosmic string. We found
in manifest  form the Green function and heat kernel of the
squared Dirac operator. We showed that the trace of the first heat
kernel coefficient is minus two times the corresponding coefficient
for the scalar field in accordance with Ref. \cite{FurMie97} and
found the first  three coefficients in manifest form. Thus, the
energy-momentum  tensor of the massive spinor field was obtained
in manifest form.  In the massless case the tensor is expressed in
terms of the second  heat kernel coefficient. The mass of the
field brings an additional  parameter, the Compton length of the
Dirac particle. Due to this  parameter the energy-momentum tensor
has an Yukawa type dependence over  the distance and the energy
density of the vacuum polarization is concentrated close to the
string, in a domain such that $\rho < m^{-1}$.

 \section*{Acknowledgments}

NK is grateful to Departamento de F\'{\i}sica, Universidade
Federal da Para\'{\i}ba, Brazil for hospitality. This work was
supported in part by Conselho Nacional de Desenvolvimento
Cient\'{\i}fico e Tecnol\'ogico (CNPq), FAPESQ-Pb/CNPq(PRONEX) and
by the Russian Foundation for Basic Research Grants No.
05-02-17344, No. 05-02-39023-SFNS.

\end{document}